\documentclass[conference]{IEEEtran}
\IEEEoverridecommandlockouts
\usepackage{cite}
\usepackage{textcomp}
\usepackage{xcolor}

\usepackage{amsmath,graphicx}
\usepackage{soul}
\usepackage{calc,amsfonts,amssymb,bm,url,color,theorem,cite}
\usepackage{psfrag,float}
\usepackage{algorithm}
\usepackage{algorithmic}
\usepackage{mathtools,lipsum,cuted,multicol}
\usepackage{filecontents}
\usepackage{subcaption,epstopdf}
\usepackage{setspace}
\usepackage{bbold}
\usepackage{enumitem}
\usepackage{array}
\usepackage{todonotes}

\definecolor{orange}{RGB}{255,107,0}

\newtheorem{Fact}{Fact}

\theorembodyfont{\rmfamily}

\newcommand\bv{\ensuremath{{\bm v}}}
\newcommand\bw{\ensuremath{{\bm w}}}

\newcommand\bh{\ensuremath{{\bm h}}}

\newcommand\jj{\ensuremath{{\frak j}}}

\newcommand{\Rbb}{\mathbb{R}}
\newcommand{\Cbb}{\mathbb{C}}

\newcommand{\bI}{{\bm I}}

\newcommand\bx{\ensuremath{{\bm x}}}

\newcommand\by{\ensuremath{{\bm y}}}

\newcommand\bH{\ensuremath{{\bm H}}}

\newcommand{\setS}{\mathcal{S}}
\newcommand{\setC}{\mathcal{C}}

\newcommand\br{\ensuremath{{\bm r}}}

\def\BibTeX{{\rm B\kern-.05em{\sc i\kern-.025em b}\kern-.08em
    T\kern-.1667em\lower.7ex\hbox{E}\kern-.125emX}}
\begin{document}

\title{An Efficient Global  Algorithm for One-Bit   Maximum-Likelihood  MIMO Detection
\thanks{This work was partly supported by a General Research Fund (GRF) of Hong Kong Research Grant Council (RGC) under Project ID CUHK 14203721.}
}

\author{\IEEEauthorblockN{Cheng-Yang Yu\IEEEauthorrefmark{1},
  Mingjie Shao\IEEEauthorrefmark{2},
  Wei-Kun Chen\IEEEauthorrefmark{1},
  Ya-Feng Liu\IEEEauthorrefmark{3},
  Wing-Kin Ma\IEEEauthorrefmark{4}
 }
 \IEEEauthorblockA{\IEEEauthorrefmark{1}School of Mathematics and Statistics, Beijing Institute of Technology, Beijing, China}
 \IEEEauthorblockA{\IEEEauthorrefmark{2}School of Information Science and Engineering, Shandong University, Qingdao, China}
 \IEEEauthorblockA{\IEEEauthorrefmark{3}LSEC, ICMSEC, Academy of Mathematics and Systems Science, Chinese Academy of Sciences, Beijing, China}
 \IEEEauthorblockA{\IEEEauthorrefmark{4}Department of Electronic Engineering, The Chinese University of Hong Kong,
 Hong Kong SAR, China}
 \small{Email: \{yuchengyang,~chenweikun\}@bit.edu.cn,~mingjieshao@sdu.edu.cn,~yafliu@lsec.cc.ac.cn, wkma@ee.cuhk.edu.hk}}


\setlength{\abovedisplayskip}{0.14cm}
\setlength{\belowdisplayskip}{0.14cm}
\setlength{\jot}{0.1cm}
\maketitle

\begin{abstract}
There has been growing interest in implementing massive MIMO systems by  one-bit analog-to-digital converters (ADCs), which have the benefit of reducing   the power consumption and hardware complexity.
One-bit MIMO detection arises in such a scenario.
It aims to detect the multiuser signals from the one-bit quantized received signals in an uplink channel.
In this paper, we consider one-bit maximum-likelihood (ML) MIMO detection in massive MIMO systems, which amounts to solving a large-scale nonlinear integer programming problem.
We propose an efficient \emph{global} algorithm for solving the one-bit ML MIMO detection problem.
We first reformulate the problem as a mixed integer linear programming (MILP) problem that has a massive number of linear constraints.
The massive number of linear constraints raises  computational challenges.
To solve the MILP problem efficiently, we custom build a light-weight branch-and-bound tree search algorithm, where the linear constraints are incrementally added during the tree search procedure and only   small-size linear programming subproblems need to be solved at each iteration.
We provide simulation results to demonstrate the efficiency of the proposed method.
\end{abstract}

\begin{IEEEkeywords}
One-bit MIMO detection, maximum-likelihood, mixed integer linear programming
\end{IEEEkeywords}

\section{Introduction}

When the massive multiple-input multiple-output (MIMO) system is realized by employing dedicated radio-frequency (RF) chains, power consumption and hardware complexity can be prohibitively high.
This has become an impediment in practical implementations for 5G systems and beyond.
To resolve the above issue,  low-resolution, particularly, one-bit analog-to-digital converters (ADCs) and digital-to-analog converters (DACs) can be employed to cut down the power consumption and hardware complexity,
because the power consumption of ADCs and DACs increases exponentially with the resolution~\cite{ADC}.
Unfortunately, the use of one-bit ADCs/DACs leads to severe quantization distortion on the signals, and this calls for customized quantized signal processing methods.

In this paper, we study the uplink multiuser signal detection problem in a massive MIMO system with one-bit ADCs.
Researchers have proposed different detection methods, including linear receivers \cite{risi2014massive,li2017channel,jacobsson2017throughput}, maximum-likelihood (ML) detection \cite{choi2015quantized,choi2016near,shao2020binary} and   maximum a posteriori (MAP) detection \cite{wang2014multiuser,wen2015bayes,studer2016quantized,thoota2021variational}.
Among the existing methods, maximum-likelihood (ML) detection is an important formulation that tries to address the quantization effect \cite{choi2015quantized,choi2016near,shao2020binary,Mirfarshbafan2020,nguyen2021linear,khobahi2021lord,shao2022accelerated,plabst2018efficient,garcia2021channel,jeon2018one}.
However, the ML formulation involves a large-scale nonlinear integer programming problem, and solving it by brute-force
exhaustive search can be computationally too demanding.
Researchers have proposed a variety of approximate algorithms to strike a balance between detection performance and computational complexity, by means of relaxation and optimization \cite{choi2015quantized,choi2016near,shao2020binary,Mirfarshbafan2020}, deep learning \cite{shao2020binary,nguyen2021linear,khobahi2021lord,shao2022accelerated}, statistical inference~\cite{plabst2018efficient,garcia2021channel,shao2022accelerated} and coding theory~\cite{jeon2018one}.
Unfortunately,   there is {\it no} efficient global algorithm for one-bit ML MIMO detection in the literature.

In this paper, we propose an efficient global algorithm for one-bit ML MIMO detection.
We first transform the one-bit ML MIMO detection problem into a mixed integer linear programming (MILP) problem.
The crux lies in that the MILP {problem} has exponentially many inequality constraints (with respect to the number of users), which results in high computational complexity if it is directly solved by an MILP solver like CPLEX \cite{CPLEX2022}.
We propose an incremental optimization strategy to alleviate the high computational burden.
It starts with a relaxed MILP problem with only a selected small subset of the inequality constraints.
Then, we iteratively and incrementally add the inequality constraints into the relaxed MILP problem.
In order to develop a light-weight global algorithm, {we solve each relaxed MILP {problem} inexactly}, which is achieved by embedding the incremental optimization strategy into one branch-and-bound procedure.
In this way, the algorithm only needs to solve linear programming (LP) subproblems with significantly much smaller problem sizes (compared with the MILP reformulation of the original problem), and is computationally efficient as demonstrated by simulations.
The proposed global algorithm offers an important benchmark for performance evaluation of  existing approximate algorithms for solving the same ML problem, showing how well they perform compared to the global ML solutions.

\section{Signal Model}
Consider a multiuser multiple-input single-output (MISO) uplink transmission, where $\tilde{K}$ single-antenna users concurrently send their signals to a base station (BS) having $\tilde{N}$ antennas.
At the BS, the received signal can be modeled by
\begin{equation}\label{eq:model}
\begin{split}
  \tilde{\by} =  ~ \tilde{\bH} \tilde{\bx} +\tilde{\bv},~~
   \tilde{\br} =  ~ {\cal Q} (\tilde{\by}),
 \end{split}
\end{equation}
where $\tilde{\bx}\in \Cbb^{\tilde{K}}$ is the multiuser transmit signal vector,   drawn from the Quadrature Phase Shift Keying (QPSK) constellation $\{ \pm 1 \pm \jj \}$;
$\tilde{\bH} \in \Cbb^{\tilde{N} \times \tilde{K}}$ is the multiuser channel matrix; $\tilde{\bv}\in \Cbb^{\tilde{N}}$ is  additive complex Gaussian noise with mean $\bm 0$ and covariance matrix $\tilde{\sigma}^2 \bI$;
${\cal Q}(x): = \mbox{sgn}(\Re(x))+ \jj\cdot \mbox{sgn}(\Im(x))$ is the one-bit quantizer for both the real and imaginary parts of $x$, and \[
\mbox{sgn}(x) =\begin{cases}
                  1, & \mbox{if } x\geq 0; \\
                  -1, & \mbox{otherwise},
                \end{cases}
                \]
is the one-bit quantization function that operates on each element of its argument;
 $ \tilde{\br}\in \Cbb^{\tilde{N}}$ is the one-bit received signal.
The one-bit MIMO detection problem is to detect $\tilde{\bx}$ from the one-bit received signal $\tilde{\br}$, with given $\tilde{\bH}$.

For   convenience of presentation, we consider the following equivalent real-valued model.
Define
\begin{equation*}
  \begin{split}
    \by  =    \begin{bmatrix}
            \Re(\tilde{\by})\\
             \Im(\tilde{\by})
           \end{bmatrix} \in \Rbb^N,  \bH   =    \begin{bmatrix}
                                   \Re(\tilde{\bH})  &     -\Im(\tilde{\bH}) \\
                                  \Im(\tilde{\bH}) &   \Re(\tilde{\bH})
                                 \end{bmatrix} \in \Rbb^{N\times K},\\
    \bx   =   \begin{bmatrix}
            \Re(\tilde{\bx})\\
             \Im(\tilde{\bx})
           \end{bmatrix}\in \Rbb^K,
            \br   =   \begin{bmatrix}
            \Re(\tilde{\br})\\
             \Im(\tilde{\br})
           \end{bmatrix} \in \Rbb^N,
                 \bv   =   \begin{bmatrix}
            \Re(\tilde{\bv})\\
             \Im(\tilde{\bv})
           \end{bmatrix} \in \Rbb^N,
  \end{split}
\end{equation*}
where $N=2\tilde{N}$, $K = 2\tilde{K}$ and $\bv$ follows the standard Gaussian distribution with mean $\bm 0$ and covariance $\sigma^2 \bI$.
We   convert \eqref{eq:model} to a real-valued form
\begin{equation}\label{eq:model_real}
  \by =  \bH \bx +\bv,~\br =  \mbox{sgn} (\by),
\end{equation}
where $\bx\in \{ -1,1 \}^K$.

We consider maximum-likelihood (ML) detection \cite{choi2016near}, which can be expressed as
\begin{equation}\label{eq:ML}
  \min_{\bx\in \{ -1,1 \}^K} f(\bx): =-\sum_{i=1}^{N} \log  \Phi \left(\frac{r_i \bh_i^{\top}\bx}{\sigma} \right),
\end{equation}
where $f(\bx)$ is the negative log-likelihood function and {$\Phi(z) = \int_{-\infty}^{z} \frac{1}{\sqrt{2\pi}}e^{-t^2}\ dt$} is the cumulative distribution function of the standard Gaussian distribution.
Problem \eqref{eq:ML} is a nonlinear integer programming problem.
To globally solve \eqref{eq:ML}, exhaustive search can be applied, which examines all feasible solutions with a complexity order of ${\cal O}(2^K)$.
To the best of our knowledge,  off-the-shelf efficient  mixed integer programming solvers such as CPLEX cannot handle $\Phi(\cdot)$, which is an integral.
In the literature, there are many approximate algorithms for solving problem \eqref{eq:ML}
that seek to strike a balance between detection performance and computational complexity~\cite{choi2015quantized,choi2016near,shao2020binary,Mirfarshbafan2020,nguyen2021linear,khobahi2021lord ,plabst2018efficient,garcia2021channel,shao2022accelerated}.
In this paper, we aim to propose an efficient algorithm for globally solving problem \eqref{eq:ML}, which can serve as a benchmark to evaluate the performance of the   existing approximate algorithms.

\section{An Efficient Global Algorithm}

In this section, we propose an efficient {\it global} algorithm for solving  problem \eqref{eq:ML}.
\subsection{An MILP  Reformulation}
We first equivalently reformulate problem \eqref{eq:ML} as
\begin{equation}\label{eq:ML2}
\begin{split}
  \min_{\bx, \bw} &~ \sum_{i=1}^{N} w_i \\
  \mbox{s.t. } &~  w_i\geq  g_i(\bx), ~i=1,2,\ldots, N ,\\
  &~ \bx\in \{ -1,1 \}^K,
  \end{split}
\end{equation}
where $\bw = (w_1,w_2,\ldots, w_N)$ and
\[
    g_i(\bx): = -\log  \Phi \left(\frac{r_i \bh_i^{\top}\bx}{\sigma} \right).
\]
Note that $g_i$ is a convex function with respect to $\bx$. The following linear inequality
\begin{equation}\label{eq:convex}
    g_i(\bx) \geq g_i(\hat{\bx}) + \langle \nabla g_i(\hat{\bx}), \bx - \hat{\bx}\rangle,~ \forall  \hat{\bx} \in \{ -1,1 \}^K
\end{equation}
is valid, where
\[
    \nabla g_i(\bx) = -\frac{\phi(r_i \bh_i^{\top}\bx/\sigma)}{\Phi(r_i \bh_i^{\top}\bx/\sigma)}\frac{r_i \bh_i}{\sigma}
\]
is the gradient of $g_i$ at $\bx$, {and} $\phi(t) = \frac{1}{\sqrt{2\pi}}e^{-t^2}$ is the probability distribution function of the standard Gaussian distribution.
Then, with \eqref{eq:convex}, we reformulate problem \eqref{eq:ML2} as
\begin{subequations}\label{eq:ML3}
\begin{align}
  (\bx^{\star}, \bw^{\star}) =\arg \min_{\bx, \bw} &~ \sum_{i=1}^{N} w_i \notag\\
  \mbox{s.t. } &~  w_i\geq g_i(\hat{\bx}) + \langle\nabla g_i(\hat{\bx}),\bx - \hat{\bx}\rangle, \label{eq:lin_ineq}\\
  &~\quad i=1,2,\ldots, N, ~\forall  \hat{\bx} \in \{ -1,1 \}^K, \notag\\
  &~ \bx\in \{ -1,1 \}^K.
  \end{align}
\end{subequations}
\begin{Fact}\label{fact:eqv}
  Problems \eqref{eq:ML} and \eqref{eq:ML3} are equivalent, in the sense that they have the same optimal solution for $\bx$.
\end{Fact}
Fact~\ref{fact:eqv} can be obtained by noting that  inequality \eqref{eq:convex} is tight when $\hat{\bx} = \bx$, which establishes the equivalence between problems \eqref{eq:ML2} and \eqref{eq:ML3}. This, together with the equivalence between \eqref{eq:ML} and \eqref{eq:ML2},  leads to the desired result.

The upshot of problem \eqref{eq:ML3} is that the inequalities \eqref{eq:lin_ineq} are \emph{linear} in both $\bx$ and $\bw$.
As a result, problem \eqref{eq:ML3} is an MILP problem.
In principle, problem \eqref{eq:ML3}  can be solved by  off-the-shelf MILP solvers such as CPLEX \cite{CPLEX2022}.
However, the number of inequality constraints in \eqref{eq:lin_ineq}  is $N\cdot 2^K$, where both $N$ and $K$ can be large in massive MIMO systems, which can lead to prohibitively high computational complexity.

\subsection{An Incremental Algorithmic  Framework}
To tackle the     computational   issue, we propose to solve problem \eqref{eq:ML3}   through an incremental optimization strategy.
We define
\[
\setC = \{ (i,\hat{\bx})~| ~i=1,2,\ldots, N, ~  \hat{\bx} \in \{ -1,1 \}^K \}
\]
 and select $\setS \subseteq \setC$ as a subset of $\setC$.
We consider the following relaxation of problem \eqref{eq:ML3}:

\begin{equation}\label{eq:rel}
\begin{split}
 (\bar{\bx}, \bar{\bw}) \in &~ \arg\min_{\bx, \bw} \sum_{i=1}^{N} w_i \\
  \mbox{s.t. } &~  w_i\geq g_i(\hat{\bx}) + \langle \nabla g_i(\hat{\bx}),\bx - \hat{\bx}\rangle, ~(i,\hat{\bx}) \in \setS,\\
  &~ \bx\in \{ -1,1 \}^K.
  \end{split}
\end{equation}
We have the following result.
\begin{Fact}\label{fact:rex}
  Consider problems \eqref{eq:ML3} and \eqref{eq:rel}. The following hold.
  \begin{itemize}
  \item [a)] $\sum_{i=1}^{N}\bar{w}_i\leq \sum_{i=1}^{N} w_i^{\star}$.
\item [b)]  if $\bar{w}_i \geq g_{i}(\bar{\bx})$ holds for all $i$, then $(\bar{\bx}, \bar{\bw})$ is also optimal to problem \eqref{eq:ML3}.
    \end{itemize}
\end{Fact}
{\it Proof}:
Since problem \eqref{eq:rel} is a relaxed version of problem \eqref{eq:ML3}, it holds that $\sum_{i=1}^{N}\bar{w}_i\leq \sum_{i=1}^{N} w_i^{\star}$. This proves a).

If $\bar{w}_i \geq g_{i}(\bar{\bx})$ for all $i$, then $(\bar{\bx}, \bar{\bw})$ is a feasible solution to problem \eqref{eq:ML2}.
Thus, $\sum_{i=1}^N \bar{w}_i \geq  \sum_{i=1}^N w^{\star}_i$.
This, together with a), implies $\sum_{i=1}^N \bar{w}_i =  \sum_{i=1}^N w^{\star}_i$.
Thus, $(\bar{\bx}, \bar{\bw})$ is an optimal solution to problem \eqref{eq:ML2}, and also problem \eqref{eq:ML3}. \hfill $\blacksquare$

\medskip

Fact~\ref{fact:rex} offers a hint to the algorithmic design.
Specifically, we start from solving  problem \eqref{eq:rel} with an $\setS\subseteq \setC$.
If $\bar{w}_i \geq g_{i}(\bar{\bx})$ holds for all $i$, then $(\bar{\bx}, \bar{\bw})$ is already optimal to problem \eqref{eq:ML3}.
Otherwise, if $\bar{w}_i < g_{i}(\bar{\bx})$ for some $i$, then the constraint
\begin{equation}\label{eq:cons}
 w_i\geq g_i(\bar{\bx}) + \langle\nabla g_i(\bar{\bx}),\bx - \bar{\bx}\rangle
\end{equation}
is added into problem \eqref{eq:rel}, i.e., adding $(i, \bar{\bx})$ into $\setS$.
Then, we solve problem \eqref{eq:rel} again with the new $\setS$.
This process is repeated until $\bar{w}_i \geq g_{i}(\bar{\bx})$ holds for all $i$.
This incremental optimization framework is described in Algorithm~\ref{Alg:OTF}.

\begin{algorithm}[t!]
\caption{An Incremental Optimization Framework for Solving Problem \eqref{eq:ML3}}\label{Alg:OTF}
\begin{algorithmic}[1]
\renewcommand{\algorithmiccomment}[1]{~~//\,\texttt{#1}}
\STATE {\bf input:} Initialization $\setS \subseteq \setC$;
\REPEAT[{one iteration}]
\STATE Solve problem \eqref{eq:rel} to obtain its optimal solution $(\bar{\bx}, \bar{\bw})$;

\IF {$\bar{w}_i < g_{i}(\bar{\bx})$  for some $i$'s}
\STATE $\setS \leftarrow \setS  \cup  \{(i, \bar{\bx}) \mid  \bar{w}_i < g_{i}(\bar{\bx}),  i=1,2,\ldots, N \}$;
\ELSE
\STATE {\bf break};

\ENDIF

\UNTIL {$\bar{w}_i \geq g_{i}(\bar{\bx})$ holds for all $i$};

\STATE {\bf output}  $(\bx^{\star}, \bw^{\star}) = (\bar{\bx}, \bar{\bw})$.

\end{algorithmic}
\end{algorithm}

%

\subsection{An Efficient Branch-and-Bound Algorithm}

Algorithm~\ref{Alg:OTF} requires (possibly) solving multiple MILP problems in the form of \eqref{eq:rel}
(e.g., by branch-and-bound algorithms \cite{Achterberg2009}) and solving each MILP problem can be time-consuming. Therefore, the complexity of Algorithm 1 can still be   high (especially when the number of iterations is large).
To further reduce the computational complexity of Algorithm~\ref{Alg:OTF}, we propose to solve each {MILP problem} \eqref{eq:rel} {inexactly}, which is done by embedding the incremental optimization iterations (cf. lines 3-8 in Algorithm~\ref{Alg:OTF}) into one branch-and-bound algorithm.
Branch-and-bound algorithms are tree search methods that recursively partition the feasible region (i.e., a rooted tree) into small subregions (i.e., branches).
In particular, our proposed branch-and-bound algorithm solves the LP relaxation in {the} form of \eqref{eq:rel2} at each iteration and gradually tightens the relaxation by adding  appropriate $(i, {\hat{\bx}})$ in the set {$\setS$} and fixing more elements of {$\bx$} to be $\{-1, 1\}.$
The resulting algorithm is still an \emph{global} algorithm to problem \eqref{eq:ML3}.
Note that the proposed algorithm only needs to solve an LP problem at each iteration, which is in sharp contrast to solving {the MILP problem} \eqref{eq:rel} in Algorithm \ref{Alg:OTF}.
Below, we present the proposed algorithm in more details.

\subsubsection{Subproblems and Their LP Relaxations}

Denote $\mathcal{F}_+$ and $\mathcal{F}_-$ as some subsets of $\{1, 2, \ldots, N\}$ such that ${x}_j=1$ for $j \in \mathcal{F}_+$ and  ${x}_j=-1$ for $j \in \mathcal{F}_-$, and $\mathcal{F}_+ \cap \mathcal{F}_-=\varnothing$.
The subproblem to explore at the branch defined by $\mathcal{F}_+$ and $\mathcal{F}_-$ is given by
\begin{subequations}\label{eq:ML3r}
	\begin{align}
		\min_{\bx, \bw} &~ \sum_{i=1}^{N} w_i \notag\\
		\mbox{s.t. } &~  w_i\geq g_i(\hat{\bx}) + \langle\nabla g_i(\hat{\bx}),\bx - \hat{\bx}\rangle, ~(i,\hat{\bx}) \in \setC,\label{eq:lin_ineq1}\\
		& ~x_j = 1, ~j \in \mathcal{F}_+, ~x_j =-1, ~j \in \mathcal{F}_-,\\
		&~ \bx\in \{ -1,1 \}^K. \label{eq:bin}
	\end{align}
\end{subequations}
{Also, consider the following LP relaxation of problem \eqref{eq:ML3r}:}
\begin{subequations}\label{eq:rel2}
	\begin{align}
		\min_{\bx, \bw}& \sum_{i=1}^{N} w_i \notag\\
		\mbox{s.t. } &~  w_i\geq g_i(\hat{\bx}) + \langle\nabla g_i(\hat{\bx}),\bx - \hat{\bx}\rangle,~(i,\hat{\bx}) \in \setS,\label{eq:lin_ineq2}\\
		& ~x_j = 1, ~j \in \mathcal{F}_+, ~x_j =-1, ~j \in \mathcal{F}_-,\\
		&~ \bx\in [-1,1 ]^K, \label{eq:cont}
	\end{align}
\end{subequations}
where $\setS \subseteq \setC$.
Problem \eqref{eq:rel2} is a relaxation of problem \eqref{eq:ML3r} by replacing $\setC$ with $\setS$ and by relaxing binary variables $x_j$'s with $x_j \notin\mathcal{F}_+\cup\mathcal{F}_-$ to $[-1,1]$.
Therefore, solving the LP problem \eqref{eq:rel2} provides a lower bound for the MILP problem \eqref{eq:ML3r}.

\subsubsection{Proposed Algorithm}
Now, we present the main steps of the proposed branch-and-bound algorithm based on the LP relaxation in \eqref{eq:rel2}.
We use $(\check{\bx},\check{\bw})$ to denote the best known feasible solution that provides the smallest objective value at the current iteration  and use $U$ to denote its objective value (called the \emph{upper bound} of problem \eqref{eq:ML3}).
In addition, {we use $(\mathcal{F}_+, \mathcal{F}_-, \setS)$ to denote subproblem \eqref{eq:ML3r} where $\setS \subseteq \setC$ relates to its current LP relaxation \eqref{eq:rel2}}, and
$\mathcal{P}$ to denote the problem set of the current unprocessed subproblems.
At the beginning, we initialize $\mathcal{P}\leftarrow\{(\varnothing, \varnothing, \setS)\}$  for some $\setS \subseteq \setC$.
At each iteration, we pick a subproblem $(\mathcal{F}_+, \mathcal{F}_-, \setS)$ from $\mathcal{P}$, and solve problem \eqref{eq:rel2} to obtain its solution $(\bx_{\sf LP}, \bw_{\sf LP})  $ and objective value $f_{\sf LP} = \sum_{i=1}^N [\bw_{\sf LP}]_i $.
Then, we have the following cases:
\begin{itemize}
	\item [(1)] If $f_{\sf LP} \geq U$, then problem \eqref{eq:ML3r} cannot contain a feasible solution that provides an objective value better than $U$ (and this subproblem does not need to be explored).
	\item [(2)] If $f_{\sf LP} < U$ and $\bx_{\sf LP} \in \{-1,1\}^K$, there are two subcases.
	\begin{itemize}
		\item [(2.1)] If $[\bw_{\sf LP}]_i \geq g_i (\bx_{\sf LP} )$ for all $i=1,2,\ldots,N$,  then $(\bx_{\sf LP}, \bw_{\sf LP})$ must be an optimal solution to problem \eqref{eq:ML3r}.
We update $ (\check{\bx},\check{\bw}) \leftarrow (\bx_{\sf LP}, \bw_{\sf LP})$ and $U \leftarrow f_{\sf LP}$.
		\item [(2.2)] Otherwise, we apply the incremental optimization strategy by adding $(i, \bx_{\sf LP})$ with all $ [\bw_{\sf LP}]_i < g_{i}(\bx_{\sf LP})$ into $\setS$ to obtain a tightened problem \eqref{eq:rel2}.
	\end{itemize}
	\item [(3)] If $f_{\sf LP} < U$ and $\bx_{\sf LP} \notin \{-1,1\}^K$, then we choose an index $j$ with $-1 < [\bx_{\sf LP}]_{j} < 1$ and branch on variable $x_{j}$ by partitioning problem $(\mathcal{F}_+, \mathcal{F}_-, \setS)$  into two new subproblems $(\mathcal{F}_+ \cup \{j\}, \mathcal{F}_-, \setS)$ and $(\mathcal{F}_+ , \mathcal{F}_-\cup \{j\}, \setS)$.
	We add the two subproblems into the problem set $\mathcal{P}$.
\end{itemize}
The above process is repeated until $\mathcal{P}=\varnothing$.
The whole procedure is summarized as Algorithm \ref{Alg:OTF2}.

\begin{algorithm}[t!]
	\caption{A Global Algorithm for Solving Problem \eqref{eq:ML3}}\label{Alg:OTF2}
	\begin{algorithmic}[1]
		\renewcommand{\algorithmiccomment}[1]{~~//\,\texttt{#1}}
		\STATE {\bf input:} Initialize $\mathcal{P}=\{ (\varnothing, \varnothing, \setS) \}$ for   some $\setS \subseteq \setC$ and $U\leftarrow +\infty$.
		\WHILE{$\mathcal{P}\neq\varnothing$}
		\STATE Choose a subproblem $(\mathcal{F}_+, \mathcal{F}_-,\setS) \in \mathcal{P}$ and set $\mathcal{P} \leftarrow \mathcal{P}\backslash \{(\mathcal{F}_+, \mathcal{F}_-,\setS)\}$;
		\LOOP
		\STATE Solve the LP problem \eqref{eq:rel2} to obtain its optimal solution $(\bx_{\sf LP}, \bw_{\sf LP})$ and objective value $f_{\sf LP} = \sum_{i=1}^N [\bw_{\sf LP}]_i $;
		\IF{${f}_{\sf LP}  \geq U$}
			\STATE {\bf break};{\COMMENT{case (1)}}
		\ELSIF{$\bx_{\sf LP} \in \{-1,1\}^K$}
		\IF {$[\bw_{\sf LP}]_i \geq g_i (\bx_{\sf LP} )$ for all $i=1,2,\ldots,N$}
			\STATE Update $(\check{\bx},\check{\bw}) \leftarrow (\bx_{\sf LP}, \bw_{\sf LP})$ and $U \leftarrow {f}_{\sf LP}$;
			\STATE {\bf break}; {\COMMENT{case (2.1)}}
		\ELSE
			\STATE $\setS \leftarrow\setS \cup  \{(i, \bx_{\sf LP}) \mid  [\bw_{\sf LP}]_i < g_{i}(\bx_{\sf LP}),  i=1,2,\ldots, N \}$; \COMMENT{case (2.2)}
		\ENDIF
		\ELSE
			\STATE Choose an index $j$ such that $-1 < [\bx_{\sf LP}]_{j} < 1$;
\STATE  Add two new subproblems $(\mathcal{F}_+ \cup \{j\}, \mathcal{F}_-, \setS)$ and $ (\mathcal{F}_+ , \mathcal{F}_-\cup \{j\}, \setS)$ into $\mathcal{P}$;~ {\COMMENT{case (3)}}
			\STATE \bf break;
		\ENDIF
		\ENDLOOP
		\ENDWHILE
		\STATE {\bf output}  $(\bx^{\star}, \bw^{\star}) \leftarrow (\check{\bx},\check{\bw})$.
	\end{algorithmic}
\end{algorithm}

In lines 3 and 16 of Algorithm \ref{Alg:OTF2}, there exist different
strategies to choose a subproblem  $ (\mathcal{F}_+, \mathcal{F}_-,\setS)$ from set $\mathcal{P}$ and to choose a branching variable index $j$ \cite{Achterberg2009}.
It is worthwhile to remark that Algorithm \ref{Alg:OTF2} can be   embedded into state-of-the-art MILP solvers like CPLEX through the so-called \emph{callback routine} \cite{CPLEX2022}, which   uses the (default) fine-tune  subproblem selection and branching strategies of MILP solvers.




\section{Simulation Results}

In this section, we provide simulation results to illustrate the efficiency of the proposed global algorithm for solving the one-bit MIMO detection problem.
The simulation settings are described as follows.
The channel $\tilde{\bH}$  is generated by element-wise i.i.d. circular Gaussian distribution with mean zero and
unit variance.
The symbols $\tilde{\bx}$ are independently and identically distributed drawn from the QPSK constellation $\{ \pm 1 \pm \jj \}$.
The signal-to-noise ratio (SNR) is defined as $
    {\rm SNR} = \frac{\|  \tilde{\bH} \tilde{\bx} \|_2^2}{\| \tilde{\bv} \|_2^2}.$
In Algorithm~\ref{Alg:OTF2}, we initialize $\setS$ by setting $\hat{\bx}$ as a zero-forcing (ZF) solution, i.e., $\hat{\bx} = \mbox{sgn}({\bH}^{\dag}{\br})$ with $^\dag$ denoting the matrix pseudo-inverse.
A number of 1,000 Monte-Carlo trials were run to obtain the bit-error rates of our algorithm and the benchmarked algorithms.

\begin{figure}[t!]
\centering
\begin{subfigure}{0.7\linewidth}
\includegraphics[width=\linewidth]{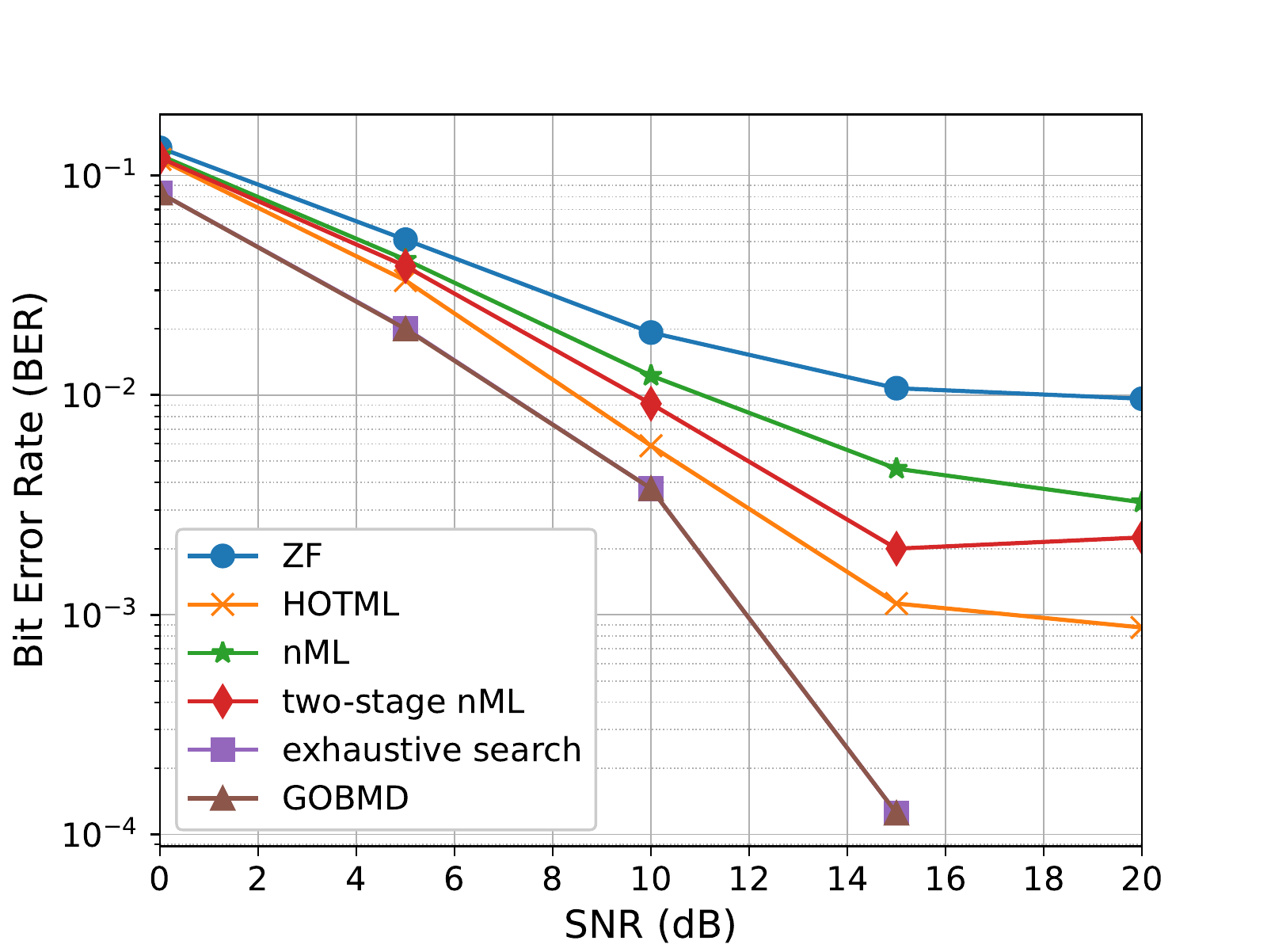}
\caption{$N=36$, $K=8$}
\end{subfigure}
\begin{subfigure}{0.7\linewidth}
\includegraphics[width=\linewidth]{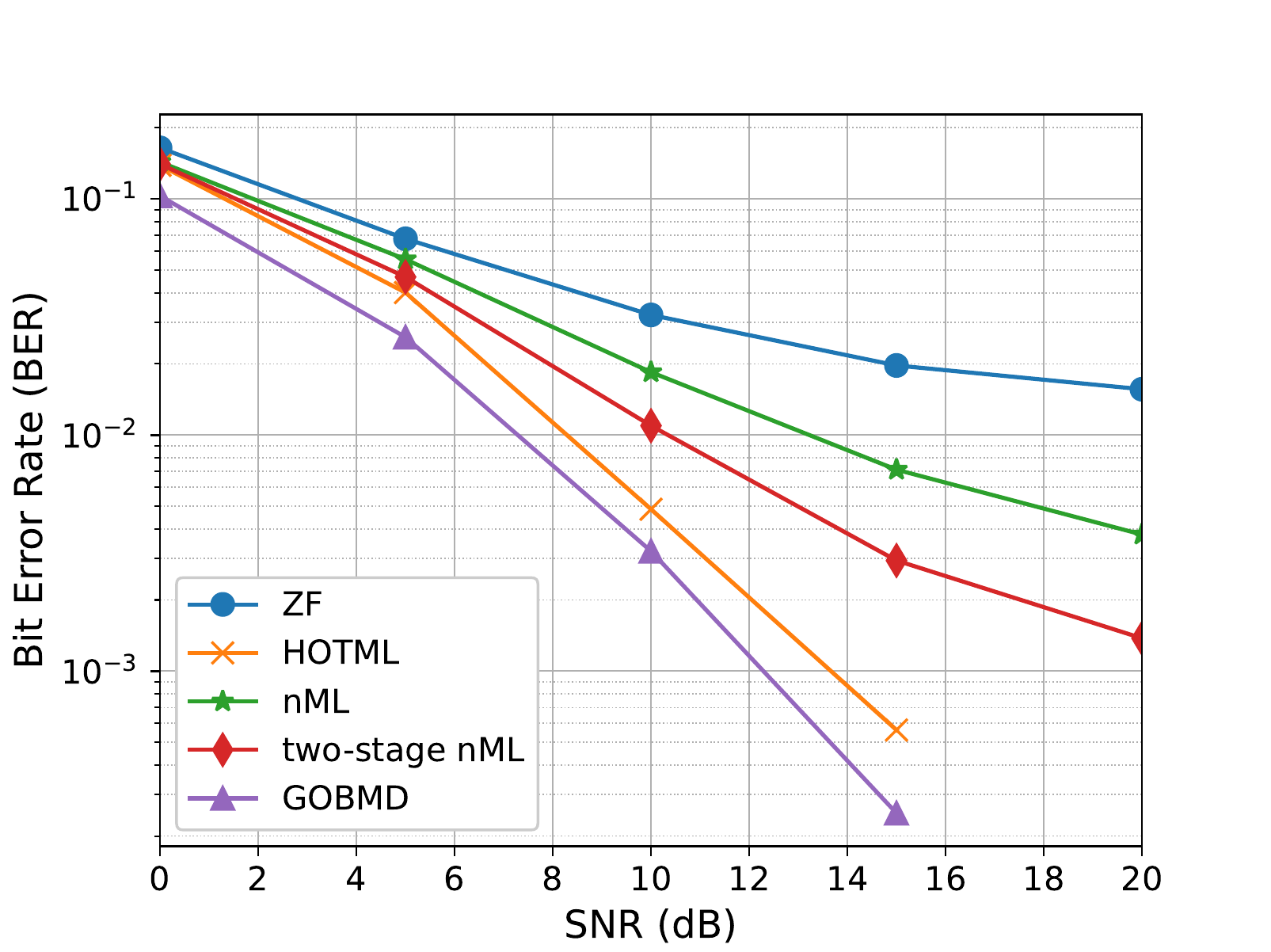}
\caption{$N=128$, $K=32$}
\end{subfigure}
\vspace*{-0.1cm}
\caption{BER performance under different problem sizes.}
\vspace*{-0.4cm}
\label{fig:BER}
\end{figure}

We first demonstrate the bit-error rate (BER) performance.
We name Algorithm~\ref{Alg:OTF2}  Global One-Bit MIMO Detection (GOBMD).
We also show state-of-the-art  algorithms that are designed to handle problem \eqref{eq:ML}, including the nML and two-stage nML in \cite{choi2016near} and HOTML in \cite{shao2020binary}.
Fig.~\ref{fig:BER} shows the BER performance under different MIMO sizes.
It is seen from Fig.~\ref{fig:BER}(a) that GOBMD achieves the same BER performance as exhaustive search, as both of them globally solve the ML problem \eqref{eq:ML}; in Fig.~\ref{fig:BER}(b),  exhaustive search is {computationally too demanding} to complete the job.
GOBMD provides an ML BER benchmark for the other algorithms for solving the same problem.

\begin{figure}[t!]
    \centering
    \includegraphics[width=0.7\linewidth]{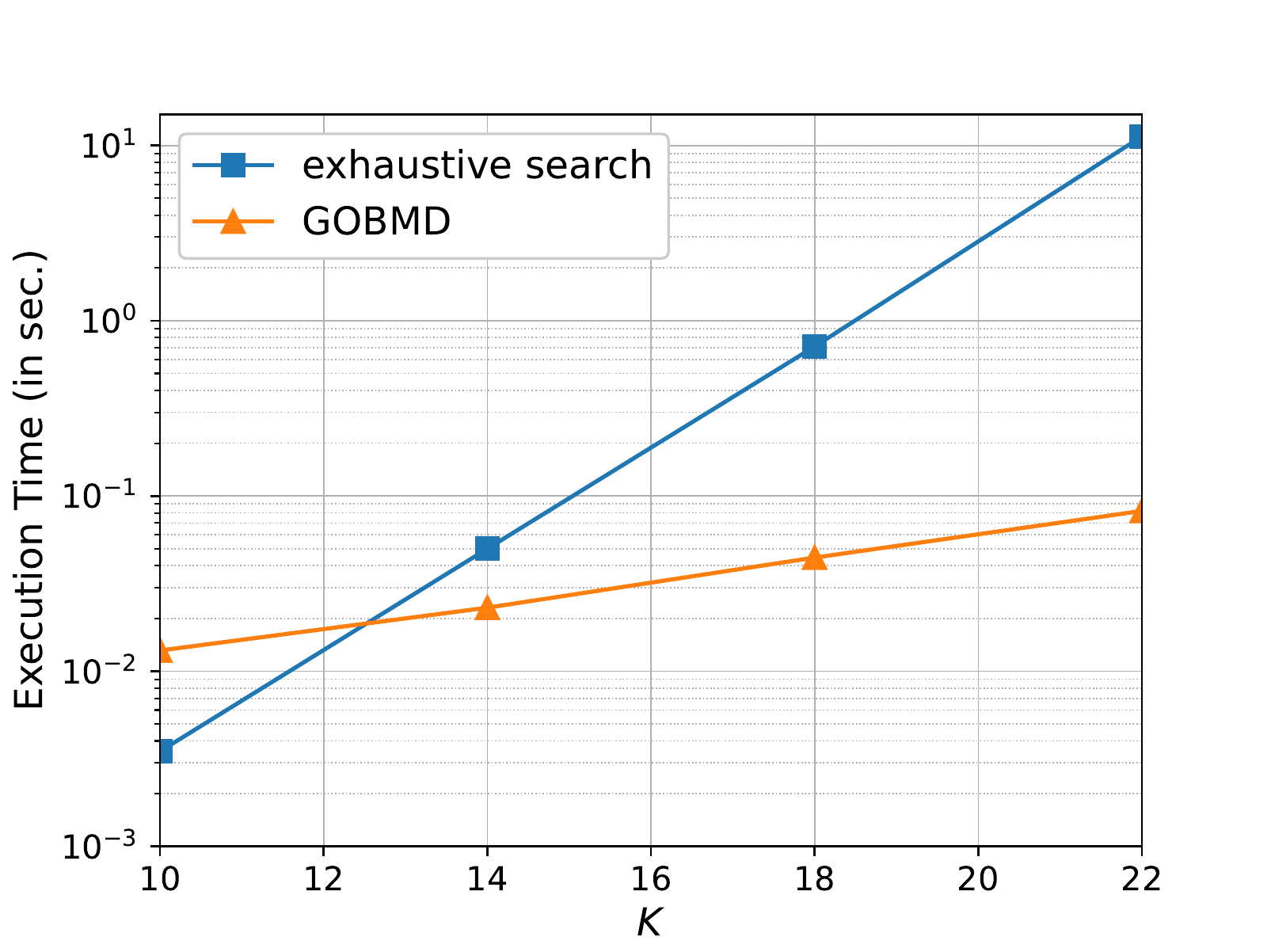}
    \vspace*{-0.1cm}
    \caption{Runtimes under different $K$; $N=36$. }
    \vspace*{-0.4cm}
    \label{fig:runtime}
\end{figure}

Fig.~\ref{fig:runtime} shows the runtime comparison between GOBMD and exhaustive search.
When the problem size is small, GOBMD and exhaustive search are computationally comparable.
However, the computational complexity of exhaustive search grows rapidly with the problem size, while that of GOBMD increases with a much slower  rate.

\begin{figure}[t!]
    \centering
    \includegraphics[width=0.7\linewidth]{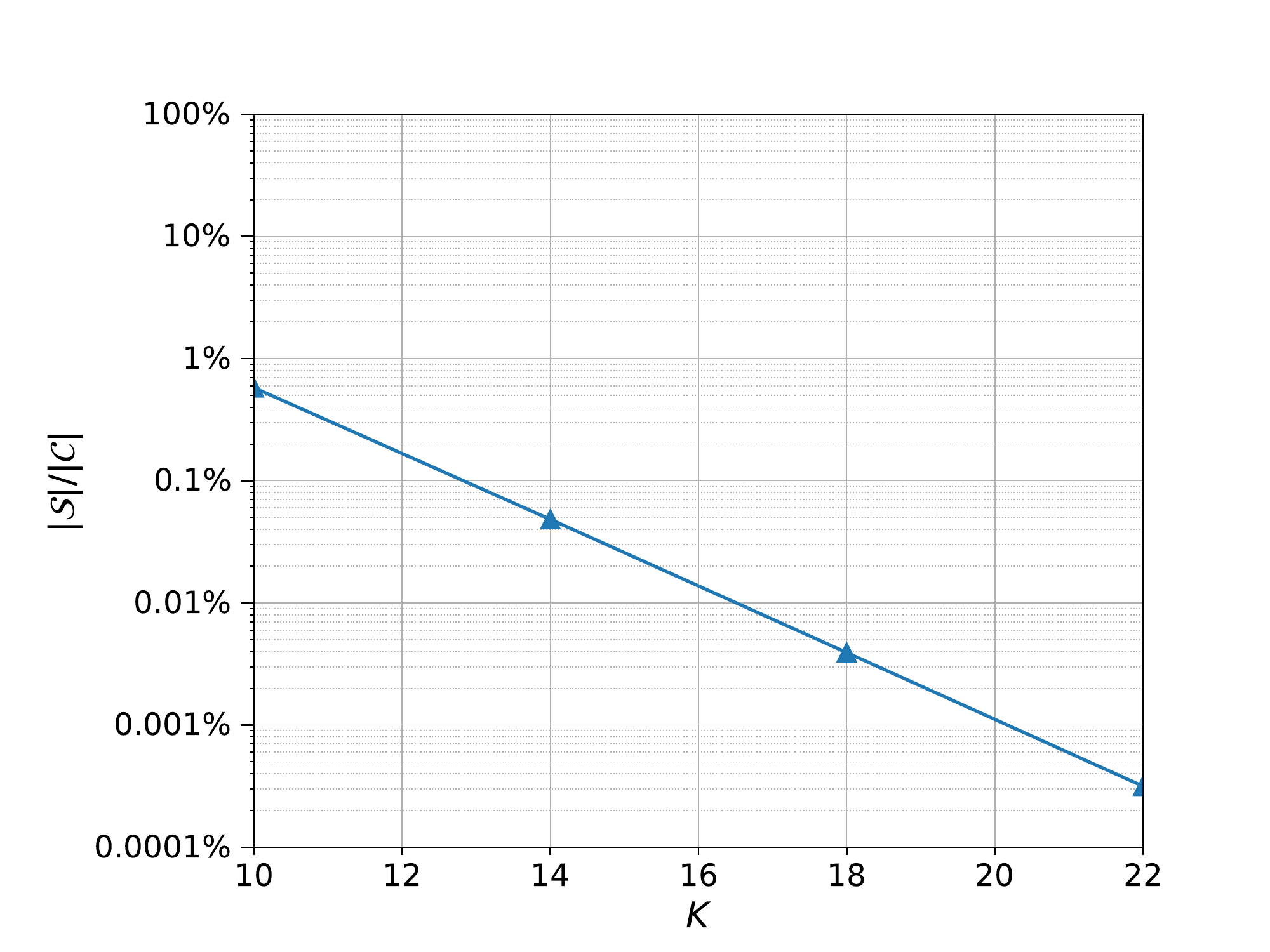}
    \vspace*{-0.1cm}
    \caption{The average ratio $|\setS|/|\setC|$ under different problem sizes. }
    \vspace*{-0.4cm}
    \label{fig:prop}
\end{figure}

Fig.~\ref{fig:prop} shows  the average ratio  $|\setS|/|\setC|$ when Algorithm~\ref{Alg:OTF2} converges under fixed $N=36$ and SNR =10 dB.
It is seen that $|\setS|/|\setC|$ is lower than 1\%.
In other words, Algorithm~\ref{Alg:OTF2} only needs to solve  LP problems that have 99\% less inequality constraints than problem \eqref{eq:ML3}.
Also, we see that the ratio  $|\setS|/|\setC|$ decreases when $K$ increases, which indicates that GOBMD has good scalability for massive systems with many users.

\begin{figure}[t!]
    \centering
    \includegraphics[width=0.8\linewidth]{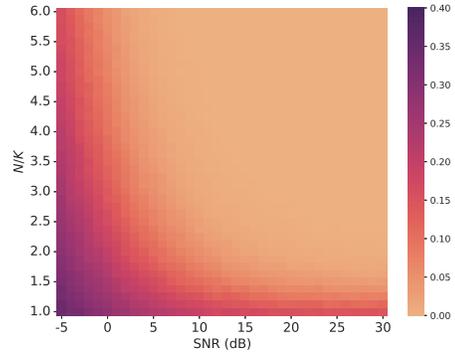}
    \caption{Phase transition of BER performance with respect to $N/K$ and SNR.}
    \label{fig:phase}
\end{figure}

Finally, as a fundamental investigation and also future work, we are interested in whether and  when the ML formulation \eqref{eq:ML} can exactly recover the user transmitted signals.
Intuitively, when the ratio between the numbers of antennas and users $N/K$ is large, and when the noise power $\sigma^2$ is small,  solving the ML detection problem will recover the user transmitted symbols (with a high probability).
The simulation result in Fig.~\ref{fig:phase} supports this intuition.
The colorbar illustrates the BER level: the darker the color, the higher the BER.
We see a clear phase transition from the left-bottom region with high BER to the right-top region with low BER.
In the future, we will quantitatively analyze  the conditions under which  the ML solution can exactly identify the user signals.
We will also extend the study to account for multi-bit quantization and higher-order modulations.


\bibliographystyle{IEEEtran}

\bibliography{ref}

\end{document}